\documentclass[journal]{IEEEtran}
\usepackage{cite}

\ifCLASSINFOpdf
\else
\fi
\usepackage{graphicx}
\usepackage{textcomp}
\usepackage{tikz-qtree}
\usetikzlibrary{trees}
\usepackage{booktabs}
\PassOptionsToPackage{hyphens}{url}
\usepackage{url}

\usepackage{pgfplots}

\usepackage{multicol}

\usepackage{multirow}

\pgfplotsset{width=4.7cm,compat=1.3}

\definecolor{mygr}{HTML}{A8A8A8}

\definecolor{mygrn}{HTML}{D3D3D3}

\usepackage{mwe}

\usepackage{scalerel}

\usepackage{balance}
\usepackage{filecontents}
\usepackage{algorithm}
\usepackage{algpseudocode}
\usepackage{listings}
\algnewcommand\algorithmicswitch{\textbf{switch}}
\algnewcommand\algorithmiccase{\textbf{case}}
\algnewcommand\algorithmicassert{\texttt{assert}}
\algnewcommand\Assert[1]{\State \algorithmicassert(#1)}%
\algdef{SE}[SWITCH]{Switch}{EndSwitch}[1]{\algorithmicswitch\ #1\ \algorithmicdo}{\algorithmicend\ \algorithmicswitch}%
\algdef{SE}[CASE]{Case}{EndCase}[1]{\algorithmiccase\ #1}{\algorithmicend\ \algorithmiccase}%
\algtext*{EndSwitch}%
\algtext*{EndCase}%
\usepackage[margin=2cm]{geometry}
\usepackage[most]{tcolorbox}
\newtcolorbox{mytextbox}[1][]{%
	sharp corners,
	enhanced,
	colback=white,
	height=6.5cm,
	attach title to upper,
	#1
}
\usepackage{pgfplots}
\usepackage{tikz}

\newcommand*\circled[1]{\tikz[baseline=(char.base)]{%
		\node[shape=circle,fill=green!15,draw,inner sep=0.1pt] (char) {#1};}}

\usepackage{enumitem}
\usepackage{xcolor}

\usepackage{amsmath}
\begin{document}
%
\title{FVCARE:Formal Verification of\\ Security Primitives in Resilient Embedded SoCs}
%
%
%

\author{Avani Dave,~\IEEEmembership{Student Member,~IEEE,}
        Nilanjan Banerjee,~\IEEEmembership{Member,~IEEE}
        and~Chintan Patel,~\IEEEmembership{Member,~IEEE}
\thanks{M. Shell was with the Department
of Electrical and Computer Engineering, Georgia Institute of Technology, Atlanta,
GA, 30332 USA e-mail: (see http://www.michaelshell.org/contact.html).}
\thanks{J. Doe and J. Doe are with Anonymous University.}
\thanks{Manuscript received April 19, 2005; revised August 26, 2015.}}

%
%

\markboth{Journal of \LaTeX\ Class Files,~Vol.~14, No.~8, August~2015}%
{Shell \MakeLowercase{\textit{et al.}}: Bare Demo of IEEEtran.cls for IEEE Journals}
%



\maketitle

\begin{abstract}

With the increased utilization, the small embedded and IoT devices have become an attractive target for sophisticated attacks that can exploit the device's security-critical information and data in malevolent activities. Secure boot and Remote Attestation (RA) techniques verifies the integrity of the device's software state at boot-time and runtime. Correct implementation and formal verification of these security primitives provide strong security guarantees and enhance user confidence. The formal verification of these security primitives is considered challenging, as it involves complex hardware-software interactions, semantics gaps and requires bit-precise reasoning.\par To address these challenges, this paper presents FVCARE - an end-to-end system co-verification framework. It also defines the security properties for resilient small embedded systems. FVCARE divides the end-to-end system co-verification problem into two modules: 1) verifying the (bit precise) initial system settings, registers, and access control policies by hardware verification techniques, and 2) verifying the system specification, security properties, and functional correctness using source-level software abstraction of the hardware. The evaluation of proposed techniques on SRACARE based systems demonstrates its efficacy in security co-verification. 
\end{abstract}

\begin{IEEEkeywords}
secure boot, formal verification, resilient system, onboard recovery, attack resilient system, small embedded systems.
\end{IEEEkeywords}

\section{Introduction}
\par The utilization of small embedded and IoT devices has increase multi-fold in recent times for collecting, processing, and transferring security-critical information and user data. It has also enabled sophisticated attackers such as \cite{furtak:2014, Stuxnet, Man1, jeep, Replay} to leak, tweek, slink or exploit the security-critical information of the device for use in malevolent activities. Denial of Service (DoS) \cite{DoS} can flood the communication interface of an application and disrupt the normal operation. Therefore, software state assurance (at run-time and boot-time) and secure communication have become essential building blocks for device security. Security primitives such as 1) secure boot measures integrity and authenticity of the software state of the device at boot-time. 2) Remote Attestation (RA) is a client-server security service, which uses a trusted third-party verifier (Vr) to send the integrity verification request to an un-trusted prover (Pr) device at runtime. The Pr computes the digest and sends the report to the Vr. Therefore, if correctly implemented, these security primitives provide a strong security guaranty about the software state of the device. Formal verification techniques are used to verify that the system posses the correct specification and security properties. \par Some of the currently available verification techniques require manual inspection or hacking skills \cite{sray:2017, sray:2015} and they can be repeated, scaled, or completely automated. Furthermore, they are to miss the bugs as manual involvement. Other existing formal verification approaches can be broadly classified in two categories: 1) Representing the firmware code in hardware by instruction level abstraction or by compiling the firmware as assembly code, and using hardware verification tools such as \cite{bs:2013, GD:2006, SM:2016} for subsequent analysis. 2) representing the abstraction of hardware as software and using software verification tools such as \cite{psob:2020,sray:2019, Mj:2017, BH:2018} to verify the necessary security properties. 
The former approach uses the complex instruction-level abstraction process that makes it ISA specific and difficult to scale. The latter approach focuses on abstracting security-specific hardware features in software.
\par Formal verification of security primitives such as secure boot and RA is considered a challenging problem, as it involves multiple complex hardware-software interactions. For example, in the case of a hybrid SRACARE based system (discussed in section~\ref{sra}), it initializes a set of hardware registers during system boot-up and applies access control policies. The verification technique needs to verify appropriate hardware registers setting (along with other firmware software features), which cannot be verified by software abstractions. Furthermore, currently available formal verification techniques use bounded model checking (BMC) \cite{} only and do not cover all system specifications, security properties edge cases. It also lacks in providing co-verification techniques of modules interaction as discussed in subsection~\ref{VC}. 
\par, Therefore, to bridge this gap, This paper presents hardware-firmware co-verification framework FVCARE. The formal co-verification process in FVCARE is compresses of 1) defining the system using a suitable mechanical model, 2) identifying and documenting the desired system properties in a succinct and intelligible way, and 3) providing proof that set system properties are satisfied. For providing proof of step (3)) FVCARE framework divides end-to-end system verification tasks into two categories: First, it uses automated hardware formal verification technique similar to that of vrased \cite{Vrased:2019}. Secondly, it uses the abstraction of hardware representation in software techniques for performing not only bounded model checking but also assertion and weakness prediction checking. It uses modular plugins with Frama-C \cite{farmac:2016} tool for software-based design specifications and security properties formal verification. 
{\bf{Research Contributions:}} The design and implementation of the proposed \emph{FVCARE} framework presents the following research contributions:
\begin{itemize}
	\item{\bf{Design Specifications \& Challenges}:} It defines the secure system design specification, Hardware (Hw), and Firmware (Fw) interactions during the authentication, secure boot, and RA computation. It also highlights the Hw-Fw co-verification challenges. 
	\item {\bf{Defines Security Properties}:} It defines the security properties for SRACARE based small embedded device.
	\item {\bf{Formal (Hw-Fw) Co-Verification Framework}:} It demonstrates the practicality of security properties specific hardware abstraction in software. It also performs formal verification using Frama-C \cite{farmac:2016} tool. Frama-C tool with three new plugins provides Weakness Prediction (WP), Value (assertions), and Linear Temporal Logic (LTL) specifications checking.  
	\item {\bf{Formal Hw Verification}:} It presents a formal hardware verification approach by converting system verilog hardware modules (for specific properties checking only)to SMV using Verilog2SMV. It verifies the specific security property using NuSMV \cite{Nusmv:2002} tool.
\end{itemize}
By combining all these, FVCARE presents the first formal co-verification framework to verify the security primitives and properties of complex firmware codes in a small embedded System on Chip (SoC) (example: SRACARE based system).  
\subsection{Organization}
Section~II covers the background, discusses the design challenges for formal verification framework, presents related work and security properties. Section~III presents targeted system design, operation, adversarial model, and scope of verification. It is followed by section~IV, covering the formal verification methodologies, discussing the hardware-software verification approaches used by FVCARE. Section~V provides the evaluation summary of the FVCARE framework by sharing verification results and findings. FVCARE evaluates the state-of-the-art system design approaches and formal verification techniques. Section~VI provides the concluding remarks for formal co-verification work of \emph{SRACARE} based SoC design.

\section{Background \& Related Work}
\par This section provides a brief overview of the background and related work of formal verification techniques.
\subsection{Background} \label{BG}
\begin{figure}[h]
	\begin{center}
		\includegraphics[width=3.2in]{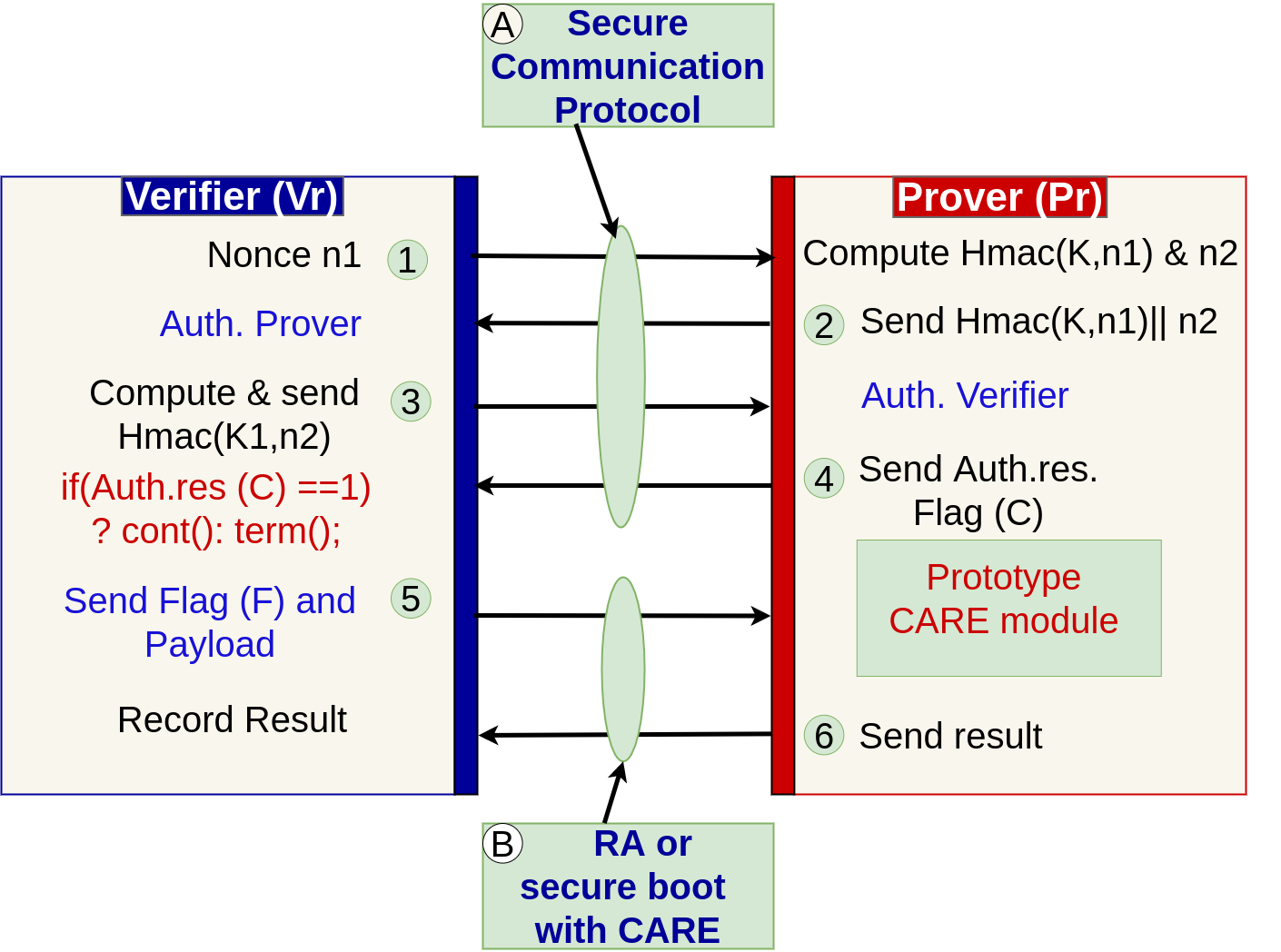}
	\end{center}
	\vspace{-2ex}
	\caption{Highlights the proposed \emph{SRACARE} system design flow. It represents the lightweight authenticated secure communication protocol and a new RA and secure boot architecture using custom \emph{CARE} module.}
	\vspace{-2ex}
	\label{fig:pro1}
\end{figure}

Although previous implementations of secure boot \cite{Tpm:2010, mcs:2015, Ope:2019, keystone, Sanctum:2018, Wong:2018, Haj:2019, NSA} and RA \cite{Vrased:2019, Seshadri2006, Perito2010}, have provided strong security guarantees about the software state of the device, the majority of them lack in providing prevention or recovery techniques, the device will be kept in hang or un-operational state upon detection of malicious code modification attacks, the device needs code reflash, which can be done by manually or over-the-air code reflash conventionally. In the event of a smart attacker corrupting the networking stack, over-the-air code reflash becomes unsuitable. Often manual code reflash becomes not feasible due to placement of the targeted devices in applications such as home security cameras, smart controllers in automotive, aviation, or industrial systems. This necessitates some form of onboard recovery techniques as represented by CARE \cite{Care:2020}. Recent work presented in SRACARE \cite{Sracare:2020} extends CARE\cite{Care:2020} by enabling RA and secure communication.  Therefore, FVCARE has selected recent SRACARE based secure RA with onboard recovery system as shown in fig~\ref{fig:pro1} for end-to-end formal verification.  
The high-level system operation can be summarized in two steps:1)~authentication of Vr and Pr devices using a secure communication protocol (steps \circled{1} to \circled{4}) and 2)~performs either remote attestation (run-time) or secure-boot (boot-time) with onboard recovery, depending on the result of step 1). Upon authentication failure, Pr sends a flag (C=0), and Vr closes the communication. Pr sends Flag (C=1) when authentication passes. Vr sends Flag (F) and payload to the Pr device to perform either secure boot with CARE or RA (as shown in steps \circled{5} and \circled{6}). The details of system design and working are covered in subsection~\S\ref{SRACAREO} and subsection~\S\ref{working}. 
\subsection{Related Work} \label{RW} 
\par Previous work presented in \cite{Ironhead:2014}, performs system-level verification by writing specification and code in dafny \cite{Dafny:2010} language, which supports automated verification using Z3 \cite{Z3:2008} SMT solver. Their tools convert dafny code to boogieX86 \cite{BoogieX86:2003} verifiable assembly language. The entire system is verified at assembly level using boogie verifier \cite{Boogie:2004}. The work presented in \cite{SBT:2017} formally verifies the UEFI secure boot system by validating PCR's content using TPM. Another co-verification approach shown by \cite{Secbootisa:2019} uses instruction-level abstraction (ISA) of hardware and applies SMACK solver to formally verify specific security properties of access control and DMA.  Recent work in \cite{psob:2020} demonstrates a framework for adversary modeling and security specification problem, followed by verification by using hyperfuzing. Another recent implementation \cite{fvap:2016} uses security properties specific hardware abstraction in software and uses SMACK solver. Vrased \cite{Vrased:2019} verifies the hardware module using Linear Temporal Logic (LTL) specifications and forces the system to reset upon security properties failure.
\par Therefore, previous research work for end-to-end security co-verification can be divided into two categories: 1) information flow analysis \cite{Balliu:2014, Sub:2016} and 2) property verification through model checking \cite{fvap:2016, Vrased:2019, ray:2019}. FVCARE belongs to the second category as it uses a software model checker to verify the security properties of complex Hw-Fw interactions. The general techniques of hardware abstraction into software and using software model checkers on the composition to verify Hw-Fw interactions are not new \cite{fvap:2016, Vrased:2019, Muk:2017}. However, the end-to-end co-verification of security properties and specification (as per subsection \ref{secprop}) for SRACARE based systems are yet to be explored.   
\par For example, in Fig~\ref{fig:pro1} the Pr device computes and checks the digest of each flash frame during the secure boot process. If the verification fails, the RE re-flashes the correct flash memory region, locks the write access, and continues the subsequent boot process. In this case, an attacker can change the recovery code's start location or redirect the system to measure boot integrity from the wrong memory region. The device requires adequate Physical Memory Protection (PMP) to prevent the write access to configuration registers and redirection of the code execution. Such scenarios require verification of hardware firmware and interaction, and any error can result in a security failure.  FVCARE focuses on concrete multi-level model checking (not just bounded) experiments along with showcasing automated hardware verification techniques, which distinguishes it from previous works.  
\section{Verification Framework Design Challenges}      \label{VC}
The scalable hardware firmware co-verification framework design faces three major challenges: 1) correct system-level abstraction, 2) definition of security properties, and 3) co-verification technique implementation.
\subsection{System Design Abstraction}
The system to be verified can be represented either hardware abstraction as software or firmware/software modules can be represented in hardware-based models. Both techniques require precisely captured sequential states of the hardware, firmware, and interacting modules. The incorrect model representation can lead to invalid verification results, a badly designed and attack-prone system. Therefore, precisely defined system security properties and boundaries for each hardware firmware components functioning are critically important for design abstraction. Section ~\ref{md} covers the available types of abstraction models and the approach used by FVCARE. 
\subsection{Security Properties Specification}
Another challenge is a system and security property specification. The hardware/firmware-based registers are setting, and component initialization, Atomicity, based temporal logic of the system can be verified by Computational Tree Logic (CTL) or Linear Temporal Logic (LTL). However, temporal logic cannot represent security properties such as controlled invocation, confidentiality, and availability. They can be verified by information flow properties analysis. Furthermore, specification of the shared system interconnect (bus) and specific SPI boundaries specification requires LTL, assertions, and weakness detection to protect the devices from \cite{Samebus:2017} attacks. A combination of system security specification tools is required for system representation and exhaustive analysis.  
\subsection{System Verification Techniques}
The co-verification of the security properties specified in either LTL, assertions, or another language needs to be checked for correctness and security assurance. This checking can be performed using Theorem Proving (TP) or Model Checking (MC). The Satisfiability Modulo Theories (SMT) solvers can be used for theorem proving. The model-checking can be used to verify the system correctness properties of finite-state transition systems \cite{FV:1993, FV:1999}. The model checking can be further classified into two types: 1) Unbounded model checking explores all reachable system transition states. 2) Bounded Model Checking (BMC) \cite{BMC:2003} restricts the search to all states reachable within the first k (bound) transitions of the system. Therefore, the selection of proper techniques becomes a crucial design component to perform exhaustive system verification. Following subsection \ref{md} covers the methodology used by FVCARE. 



%

\section{Overview of SRACARE}\label{sra}
Before going into the details of verification methodology, this section covers the summary of system design and operation of SRACARE \cite{Sracare:2020} based system used for formal verification in proposed FVCARE.
\subsection{{System Design}} \label{SRACAREO}
\par
\begin{figure}[h]
	\begin{center}
		\includegraphics[width=3.4in]{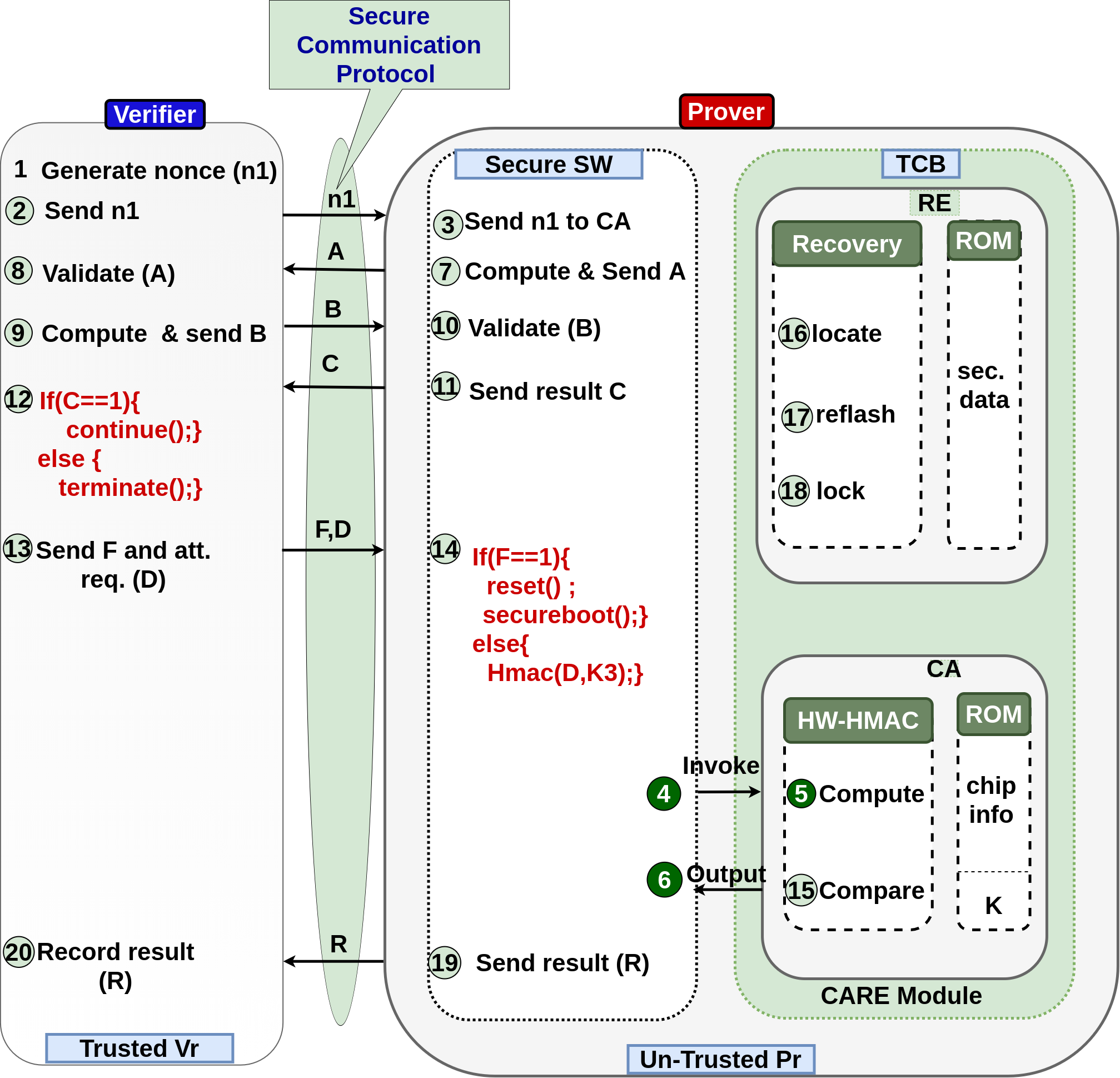}
	\end{center}
	\caption{Highlights the system design and key contributions of \emph{SRACARE}: 1) Novel lightweight, secure authenticated communication protocol (steps {1} to {12}), and 2) Secure boot with \emph{CARE} and remote attestation architecture for the P\textsubscript{r} device (steps {13} to {20})).
		\vspace{-1em}
		\label{fig:main}}                          
\end{figure}

 \emph{SRACARE} system's top-level design overview is presented in Fig~\ref{fig:main}.
The core security enhancing features of SRACARE based system are:   1) It implements lightweight, secure communication protocol and 2) It demonstrates the lightweight implementation of secure boot system with on board recovery engine (by using \emph{CARE} module). It also implements sample RA architecture for run-time software state assurance of the Pr device. The notations and definitions used for the communication are listed in Table~{\ref{reftab}}. The detailed working of the secure communication protocol(steps (1) to (12) from Fig~\ref{fig:main}) is covered in subsection~\S\ref{SCP}. 
\begin{table}[h]\caption{Notations and description \label{reftab}}
	\begin{tabular}{ |p{2cm}|p{5.6cm}|  } 
		\hline
		Notation  & Description \\
		\hline       
		\hline
		n1   &V\textsubscript{r}'s nonce for freshness \\
		n2   &P\textsubscript{r}'s nonce for freshness \\
		& n2 = Hmac(K, T) \\
		& T = hash(CHIP INFO.) $\oplus$ n1 \\
		K &Symmetric key for HMAC \\
		Hmac(K, m) &H(($K'\oplus 0$x$5C5C$) $||$ H(($K'\oplus 0$x$3636$) $||$ m)) \\
		A &A = Hmac(K, n1) $>>$ n2  \\
		B &B = Hmac(K\textsubscript{1}, n2)   \\
		C &C is a true or false result of the validation of B. \\
		D &D consists of parameters S\textsubscript{addr} and L as payload for attestation \\
		F &Reset Flag  \\
		S\textsubscript{addr} &Start address of flash memory for hashing \\
		L &Lenth of the memory region to be hashed \\
		R &Final Result  \\
		$K'$ &$\left\{
		\begin{array}{ll}
			H(K) \;\;K\;is\;larger\; than \; the \;block \;size  \\
			K \;\;\;\;\;\;\; otherwise\\
		\end{array}
		\right.$ \\
		m &Memory region to be attested, derived from S\textsubscript{addr}, L \\ 
		H &Cryptographic hash function \\    
		$K'$ &Key derived from the secret key K   \\
		K\textsubscript{1} &K\textsubscript{1}= (Hmac(K, n1) $\oplus$  n1 $\oplus$ n2))   \\
		$||$ &Denotes concatenation \\    
		$\oplus$ &Denotes bitwise exclusive or (XOR)  \\  
		CA &Code Authentication \\    
		RE &Resilience Engine \\  
		RA &Remote Attestation  \\  
		\hline
	\end{tabular}
\end{table}
The proposed secure communication protocol has two advantages over conventional authenticated communication protocols: (1) It authenticates both end devices (the P\textsubscript{r} and V\textsubscript{r}) in the communication and provides resilience from \cite{Man1}, \cite{Replay}, and \cite{DoS} attacks. (2) It does not require additional computationally heavy system resources such as TRNG, Authenticated Encryption with Associated Data (AEAD), Elliptic Curve Digital Signature Algorithm (ECDSA) or complex Message Authentication Code (MAC) to satisfy A3 security properties listed in section~\S\ref{secprop}. 
\begin{figure}[h]
	\begin{center}
		\vspace{-1em}
		\includegraphics[width=3.3in]{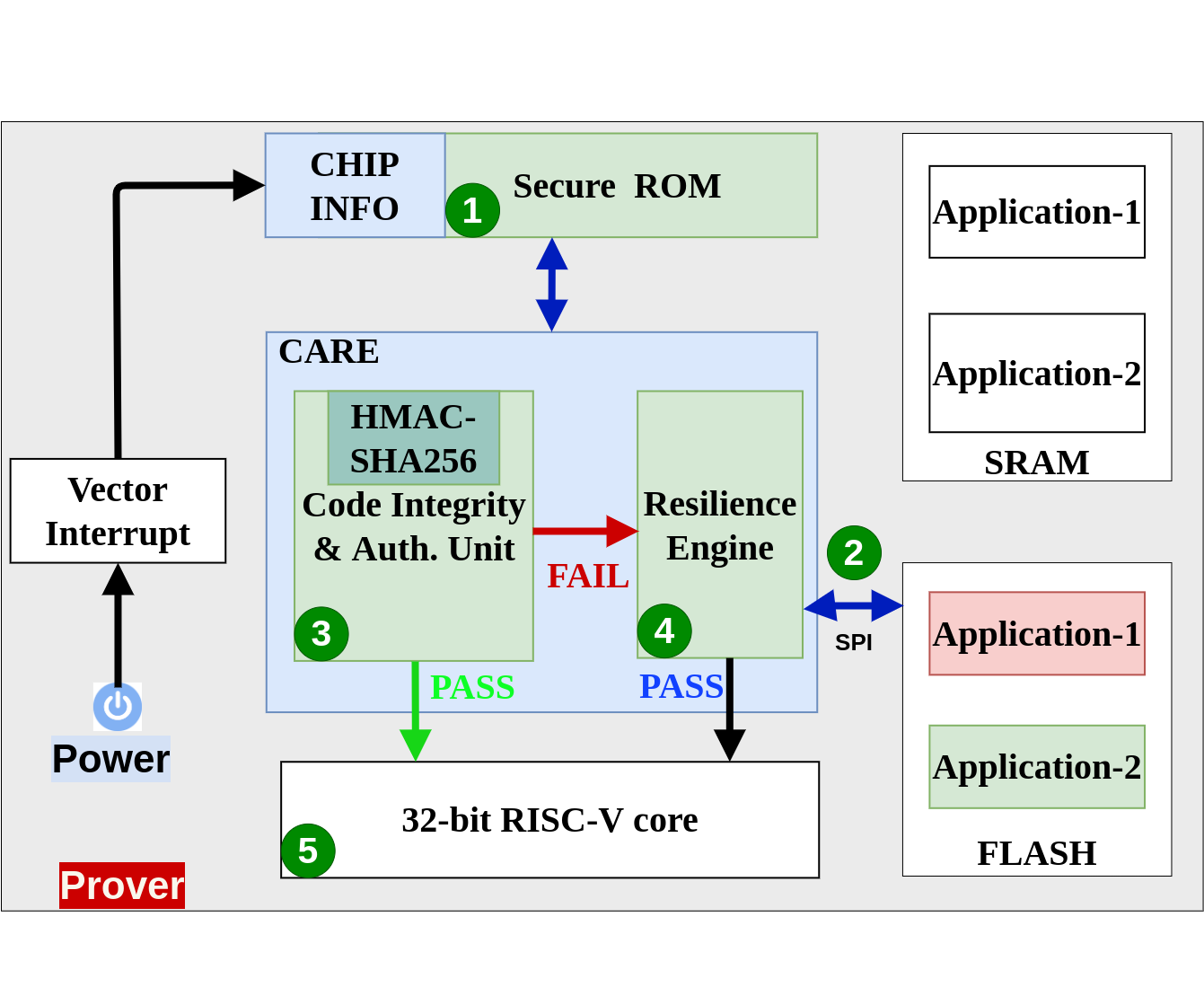}
	\end{center}
	\vspace{-3.5ex}
	\caption{Shows the architecture design of \emph{SRACARE} based P\textsubscript{r} system, highlighted are the key design modules. The pass arrows indicate that only the known good code will be allowed to be executed on the \mbox{RISC-V} core at any given time.}
	\vspace{-2.0ex}
	\label{fig:care}
\end{figure} 
Fig~3 shows the internal architecture design of Pr device to satisfy the security properties from A1, A2, A4 to A12 from subsection~\S\ref{secprop}. \emph{SRACARE} based P\textsubscript{r} system follows design choices {\includegraphics[width=.3cm,height=.3cm]{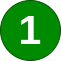}} to {\includegraphics[width=.3cm,height=.3cm]{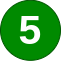}}, as highlighted in Fig~3. The P\textsubscript{r} performs either the RA or secure boot with \emph{CARE} by following steps {13} to {20} from Fig~2. The detailed working of the system is covered in section~\S\ref{working}.
\subsection{{System Operation}} \label{working}
\par
\begin{figure}[h]
	\begin{center}
		\includegraphics[width=3.3in]{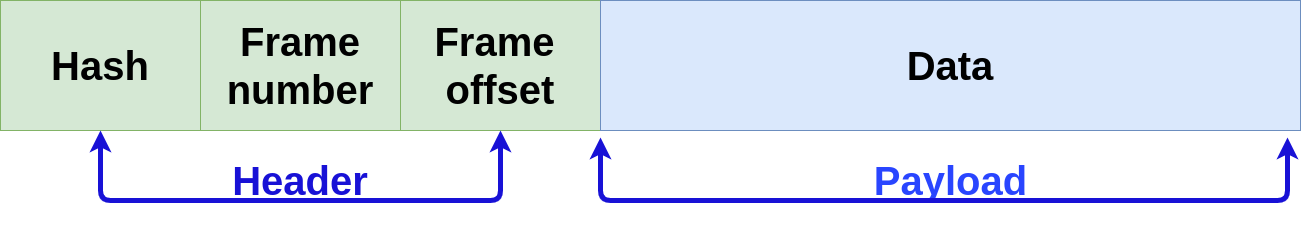}
	\end{center}
	\vspace{-3ex}
	\caption{Represents the frame data structure. The header contains the digest of the entire frame, frame number, and flash offset location. The payload contains corresponding data for each frame.  
		\vspace{-2ex}
		\label{fig:frame}}
\end{figure}

 The system operation of \emph{SRACARE} based system is divided into four main steps: 1) Secure Communication Protocol, 2) Secure Boot, 3) Resilience and Recovery, and 4) Remote Attestation. \\ 
{\bf{1) Secure Communication Protocol}:}\label{SCP} The secure communication starts when the Vr sends nonce n1 to the Pr device. The un-trusted Pr device uses novel n2 generation techniques by computing Hmac(K, T).
\[n2 = Hmac(K, T)  \]     
\begin{equation}\label{eq23}
	T = hash(CI\textsubscript{start}, 16)\;\; xor \;\;n1
\end{equation}
Where T is computed by xoring the digest of first 16 Bytes of the chip info memory and n1. The term CI\textsubscript{start} indicates the starting location of Chip Info (CI) memory. The P\textsubscript{r} generates A = (Hmac(K, n1) $>>$ n2) by appending n2 with Hmac(K,n1) and sends it to the V\textsubscript{r}. The V\textsubscript{r} validates the authenticity of the P\textsubscript{r} by recomputing Hmac(K, n1) and matching it with the received value. The V\textsubscript{r} derives the new secret key K\textsubscript{1}, computes Hmac(K\textsubscript{1}, n2), and sends the result to the P\textsubscript{r}. The P\textsubscript{r} follows the appropriate generation and validation steps to authenticate the V\textsubscript{r} and sends the result Flag C (step {11} from Fig~2) to the V\textsubscript{r}. \emph{SRACARE} closes the POC UART connection (it can be Xbee or other) between the P\textsubscript{r} and V\textsubscript{r} devices when the V\textsubscript{r} receives (C==0) (in step {12} from Fig~2), else it sends the Flag F defining the next action and associated payload to the P\textsubscript{r}. \\ {\bf{2) Secure Boot}:} If the received Flag (F) is set (F==1), then the P\textsubscript{r} calls system reset function and performs the secure boot with \emph{CARE}. Note that steps {\color{blue}4} to {\color{blue}6} in Fig~\ref{fig:main} are represented differently to denote that those steps will be part of both RA or secure boot. However, the sequence of execution will be different. As depicted in Fig~3, the secure boot sequence starts with the system power-on. It locates and executes the First Stage Boot Loader (FSBL) code from secure ROM to initialize the SPI and flash controllers, read chip information such as - device UUID, board version, symmetric share key, and hand off the control to the second stage boot code called the bootstrap. The bootstrapping process divides the flash image into a 1~KB frame chunks and sends it one at a time to the host via SPI bus for integrity and authenticity check. Each frame consists of the header and associated payload, as indicated in Fig~{\ref{fig:frame}}.
The header section of the data frame contains the digest of the entire frame, frame number, and the flash offset location. The offset location is the flash memory offset location used for the frame reflashing. The payload contains the corresponding data for each frame. This work has leveraged the Hash-based Message Authentication Code's (HMAC) feature for signing (authenticating) the data and the SHA256 feature for integrity check for each frame to reduce hardware footprint and cost. Secure boot follows steps {\color{blue}4}-{\color{blue}5}-{15}-{16}-{17}-{18}-{\color{blue}6} from Fig~2 for each frame, and upon digest mismatch detection, the P\textsubscript{r} triggers the RE else the device will continue the normal boot process.\\ {\bf{3) Resilience Engine}:} RE follows steps {16}-{17}-{18} from Fig~2 to locate the affected memory region, reflashes the corrupted flash memory region with the known good software code from secure ROM, and locks the unauthorized access to the flash region using Physical Memory Protection (PMP) mechanism of the \mbox{RISC-V} processor.\\ {\bf{4) Remote Attestation}:} If the received Flag (F==0) value is not set, the P\textsubscript{r} performs remote attestation based on the payload provided by the V\textsubscript{r}, which consists of the start location and the length of the information to be attested. The P\textsubscript{r} follows steps {\color{blue}4}-{\color{blue}5}-{\color{blue}6} sequence from Fig~2 to compute the digest and it sends the report to the V\textsubscript{r} (steps {19} and {20} from Fig~2).


\subsection{{Security Properties}} \label{secprop}
FVCARE has identified twelve (A1- A12) security properties for targeted SRACARE \cite{Sracare:2020} based system with secure boot, RA, and onboard recovery needs to satisfy for end-to-end co-verification. They can be broadly classified into four domains:  1) System Initialization \& Secure Communication, 2)Key Protection, 3)Safe Execution, and 4)Safe Recovery.\\
{\bf{1) System Initialization \& Secure Communication}:} This subsection defines required security properties during the startup of the system \& peripherals initialization. It also focuses on defining security properties for derived keys generation and device authentication at the Pr side.
{\bf{A1.~Start-up Checking}:} The startup security properties require the correct implementation of start addresses and range of ROM, RAM, flash memory regions, and MMIO device mapping, based on the system's design specification.
{\bf{A2.~Peripheral Initialization}:} This security property includes initialization of system registers, flash controller, SPI, and baud rate setting of UART. It involves the function calls from firmware to initialize the respective hardware modules (UART, SPI) and registers.
{\bf{A3.~Secure Communication}:} The design under test uses a secure device authentication protocol for Pr \& Vr devices authentication. As discussed in subsection~\ref{working}, it uses novel derived key and nonce generation techniques. This security property requires proper derived key and nonce generation techniques.
{\bf{2) Key Protection}:} This property ensures that all the secure device information, crypto, and derived keys are stored in secure ROM regions and protected by access control policies. {\bf{A4.~Key Confidentiality}:} This security property validates that the secure key (K) and derived K' are stored in a protected ROM memory region. {\bf{A5.~Access Control Enforcement}:} It defines the PMP access control policies to protect the system from unauthorized memory accesses. It involves both hardware and software system modules.\\
{\bf{3) Safe Execution}:} This property ensures the correct implementation, controlled invocation, and un-interrupted execution of the secure boot code.
{\bf{A6.~Functional Correctness}:} This security property requires the implementation and functional correctness of hardware-based crypto-core. {\bf{A7.~Atomicity}:} This specification ensures that once triggered; the code execution should not be interrupted. {\bf{A8.~Error Free Execution}:} All the hardware (IPs) and software sub-modules should have error-free execution. {\bf{A9.~Controlled Invocation}:} The security property defines no interrupt execution, DMA operations, and debugger usage are allowed during the secure boot process.\\
{\bf{4) Safe Recovery}:} The correct implementation and error-free execution of RE sub-module code execution for recovery. 
of 
{\bf{A10.~ Attack Detection}:} This security property ensures that the corrupted flash memory region is identified correctly during the secure boot process. 
{\bf{A11.~Secure Reflash}:} This property defines ensures that the affected device is re-flashed with the appropriate recovery code (from secure ROM). It requires validation of the start address and size of the flash memory and recovery data. 
{\bf{A12.~Access Controls}:} This security property ensures that proper access control policies are applied after recovery code reflash from secure ROM. It protects the device from future flash modification attacks.\\
\subsection{{Scope of Verification}}
\par This work provides a formal co-verification framework for SRACARE based system with secure boot, RA, and onboard recovery. All system specification, security properties, and the hardware-software modules setting \& interactions are formally verified. However, formal verification of the processor is out of scope for this work. The hardware system model (crypto-core and other TCB modules) are represented in Register Transfer Level (RTL), and software - boot process, bootstrapping, and resilience engine codes are written in C programming language. The functionality and security properties are specified in Linear Temporal Logic (LTL) and assertions. The model checker NuSMV \cite{Nusmv:2002} is used for hardware verification. The RTL to SMV model specifications are generated using Verilog2SMV \cite{auto:2016} tool. For the formal software verification, Frama-C \cite{farmac:2016} tool with three different plugins was used for functional, specification, weakness prediction, and LTL model checking. 


\section{Formal Verification Approach}
This section covers the FVCARE's approach for end-to-end system modeling, abstraction, and formal verification. 
\subsection{\bf{Formal Co-Verification Methodology}}\label{md}
As discussed earlier, the end-to-end security properties verification of a system with secure boot, RA, and recovery engine becomes a complex problem, and we argue that only hardware verification or verification of only software abstraction of hardware can miss out on critical security bugs during the security properties setting or interaction. Therefore, the FVCARE framework proposed two-step end-to-end security properties verification for the system under test. In the first step of the verification, 1)it verifies the security-critical hardware components of the FVCARE system using hardware verification technique and 2) for the overall system and software verification, it uses source level (C) abstraction of hardware technique.  
\begin{figure}[h]
	\begin{center}
		\includegraphics[width=3.3in]{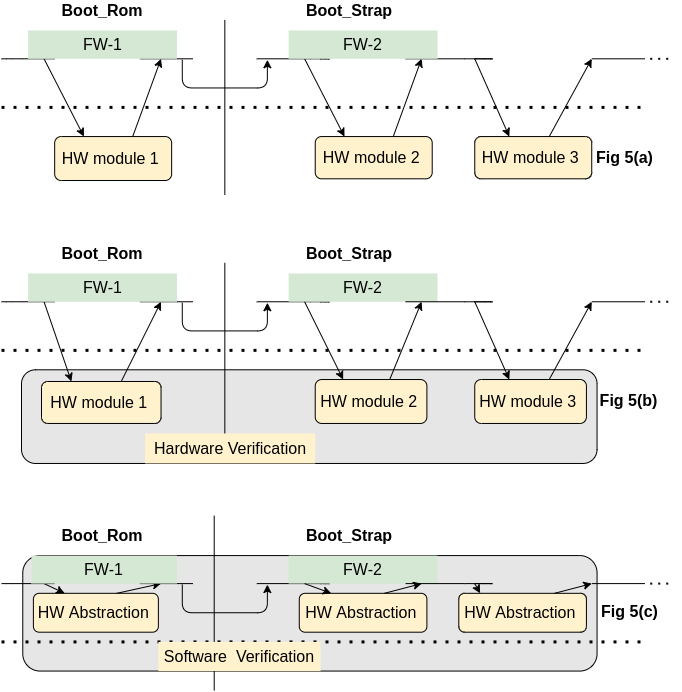}
	\end{center}
	\caption{Formal Verification Methodology 
		\label{fig:fv1}}
\end{figure}    
\par A pictorial overview of the verification methodology is shown in Fig~\ref{fig:fv1}. The goal is to co-verify the firmware, software, and interacting hardware components.  Fig~5(a) shows firmware verification is a complex task as it involves multiple interactions between software and hardware modules. In the first step, Boot\_Rom firmware code (FW-1) initializes the UART, internal registers with boot settings (HW module 1), and transfers the control to second stage Boot\_Strap firmware code (FW-2). The FW-2 code initializes the SPI, applies the access control policies (HW module 2), and performs a chain of integrity \& authenticity measurements using hardware crypto-core (HW module 3). The hardware verification method is shown in Fig5(b) and covered in subsection~\ref{hwfv}. The framework uses the source level abstraction of the hardware module for end-to-end security properties verification by software abstraction approach. Software verification is presented by Fig~5(c) and covered in subsection~\ref{swfv}.
\subsection{\bf{Hardware Verification Technique}}\label{hwfv}
\par The framework formally verifies the security-critical hardware component - crypto-engine HMAC-SHA256 's security properties and functional correctness. It also verifies the access control policies and internal registers settings for UART's baud rate, SPI's initialization, and the flash controller. The formal verification of all other hardware modules, including the processor, is out of this work scope. 
\begin{figure}[h]
	\begin{center}
		\includegraphics[width=3.3in]{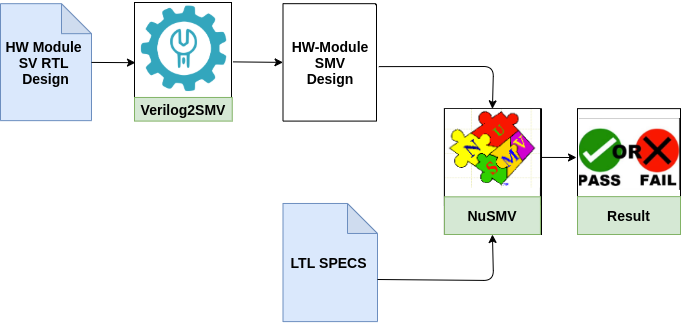}
	\end{center}
	\caption{Top-level design of hardware verification framework. The RTL design is converted to SMV using verilog2SMV tool. The hardware specifications and security properties represented in LTL logic are verified using NuSMV (SMT solver) and pass or fail results are generated. 
		\label{fig:hwfv}}
\end{figure}  
\par The hardware module listed above are written in system Verilog. Linear Temporal Logic (LTL) is used to formalize the system specifications, security properties (A1-A6) from subsection~\ref{secprop}, and invariant that should hold throughout boot code execution. FVCARE automates the system Verilog to SMV conversion process by using Verilog2SMV \cite{auto:2016} tool. It then uses the NuSMV \cite{Nusmv:2002} model checker to verify the correctness of LTL specifications. Upon failure of the formal verification, the hardware module was redesigned for security assurance.
\subsection{\bf{Software Verification Technique}} \label{swfv}
\par For end-to-end system modeling Instruction Level Abstraction (ILA) of the software of the system. ILA approaches \cite{bs:2013, GD:2006, SM:2016} involves the deep understanding and cycle-accurate conversion of the software system into instructions for that processor, and it cannot scale for fast pace changing markets. Another approach uses the hardware modules' abstraction into suitable software code (C programs), as shown in Fig~5(c). After source-level abstraction, FVCARE uses Frama-C \cite{farmac:2016} (FRAmework for Modular Analysis of C code) for software verification tool. \par The Frama-c tool can be used for buffer-overflow, pointer safety, exceptions, termination, K-induction, and invariant checking. It can also be used for specific system properties checking using assertions and LTL specifications. The Frama-c framework has a collection of interoperable, scalable, and sound software analysis tools. 
In this work, the Frama-c framework is used with three main plugins: 1) Abstract interpolation-based Value plugin 2) Weak prediction (WP) for deductive verification, and 3) System specification \& security properties verification using LTL specifications. 

1) The abstract interpretation framework based on the VALUE plugin is used to compute the over-approximations of possible values of program variables at each program end-point. It uses formal behavioral specification language ACSL (ANSI/ISO C Specification Language) to specify the C program's functional properties and contracts. ACSL uses clauses for precondition verification, ensures clauses for post-condition verification, assign for global variables, and loop invariant clauses are used for loop iterations. 



\begin{figure}[h]
	\begin{center}
		\vspace{-1em}
		\includegraphics[width=3.3in]{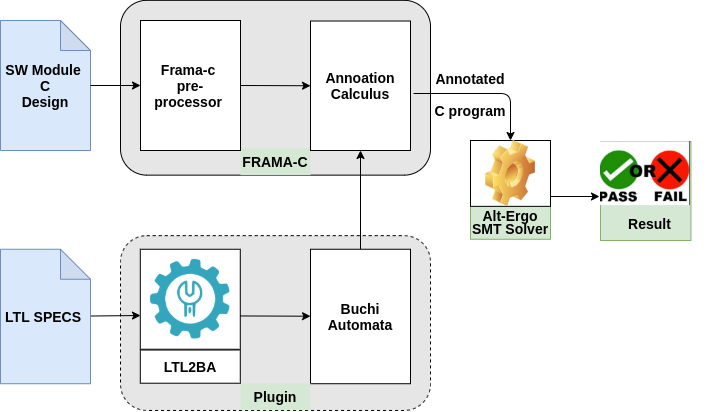}
	\end{center}
	\caption{Presents the architecture design of the proposed framework. The pass arrows indicate that only the known good code will be passed to the \mbox{RISC-V} processor core for execution in any given case.
		\label{fig:swfv}}	
\end{figure}

2) The Weak Prediction (WP) plugin verifies that the given C code satisfies its specification expressed as ACSL annotations. The weak prediction provided by \cite{dikstra:1975} reduces any deductive verification problem to establishing the validity of first-order formulas called verification conditions. FVCARE framework then uses Alt-Ergo SMT solver to prove the verification conditions generated by WP. \\
3) The Aorai plugin is used to verify the system properties represented in LTL specifications. The software verification framework setup with Frama-c and Aorai plugin is shown in Fig~\ref{fig:swfv}.
\label{design}
 \\
The Frama-c pre-processor module converts the C program into annotation calculus. The LTL2BA tool converts the  LTL specification into Buchi automaton, combined with annotation calculus to generate an annotated C program. Alt-Ergo SMT solver is used to prove the verification conditions with fail or pass results formally. The functions and code were modified to satisfy the security properties specified in A1 to A12.

\section{\bf{Evaluation}}
This section covers details of verification techniques selection for each of the security components in SRACARE based design under test. It shows verification results and timing analysis. Finally, it compares the proposed FVCARE framework with state-of-the-art secure boot, RA, and formally verified systems. Based on the security properties specification in subsection~\ref{secprop} and system operation in subsection~\ref{working}, the end-to-end system verification is divided into the following subtasks. 
\subsection{\bf{Verification of System Initialization}}\label{vfss}
\par The formal verification of start addresses and range of ROM, RAM, flash memory regions, MMIO device mapping (security property A1), and registers initialization (A2) is performed by using the hardware verification technique \ref{hwfv}. The peripheral initialization for security property A2 is verified using co-verification techniques.   
\subsection{\bf{Verification of Secure Communication}}\label{vfss}
To limit the secure formal communication verification problem, the FVCARE focuses only on verifying the Pr side security properties. Therefore, this work's verification efforts assume that nonce n1 and Flag selection values are provided to the prover (Pr) device. The LTL model was designed based on the system specifications to validate novel nonce generation techniques, and assertions were passed.
\subsection{\bf{Verification of Secure Boot}} \label{vfsb}
The secure boot verification process involves multiple transactions between Hw-fw. Therefore, it uses software abstraction of the hardware. The verification of security properties A3 to A9 was also included in this work. The security properties were passed in LTL formulas, assertions, and annotations in Frama-C. The hardware verification of crypto-core (HMAC-SHA256) is covered in subsection~\ref{PA2}.
\subsection{\bf{Verification of Remote Attestation}}\label{vfra}
\par To reduce the complexity of verifying RA communication protocol, FVCARE checks the preset flag condition (F=1 from Fig~2), start location, and the flash region size. The annotations validation, LTL specification checking for digest computation is performed using Frama-C. Formal verification of 1~KB of the digest computation requires 0.02s on intel NUC i-7 8th generation running @ 3.4GHz. 
\subsection{\bf{Verification of Resilience Engine}}\label{vfre}
\par Since the RE module was implemented in software, the verification is also performed using the software-based Frama-C tool with plugins. The important verification parts in this are to validate frame numbers and flash memory locations. The framework also verifies proper access locks during and after the code reflash.   
\subsection{\bf{Verification of Security properties}}\label{PA2}
\par The security properties specification,  formal verification approach (Hw or Fw based), execution time, and results are represented in Table~\ref{vft}. It also provides details about hardware or firmware-based formal verification techniques usage for set property. Selected set of the security properties, their verification technique (Hw/Fw), the time required on Intel Next Unit of Computing (NUC) i7 @3.4Ghz. \begin{table*}[ht]
	\caption{Qualitative comparison between secure boot/RA techniques targeting lightweight embedded devices.}
	\centering
	\begin{tabular}  {@{}lcccccccc@{}}
		\toprule
		\multicolumn{1}{l|}{Parameters}        & \multicolumn{1}{c|}{\cite{Sracare:2020}, \cite{Care:2020}} & \multicolumn{1}{c|}{Healed} & \multicolumn{1}{c|}{Ref.\cite{Secerase:2010}} & \multicolumn{1}{c|}{Ref.\cite{Haj:2019}} & \multicolumn{1}{c|}{Sanctum} & \multicolumn{1}{c|}{Ref.\cite{fvap:2016}} & \multicolumn{1}{c|}{Ref.\cite{Vrased:2019}} & \multicolumn{1}{c}{Ref.\cite{ray:2019}} \\ \midrule
		\multicolumn{1}{l|}{Design Type}         & \multicolumn{1}{c|}{Hybrid}  & \multicolumn{1}{c|}{SW} & \multicolumn{1}{c|}{Hybrid} & \multicolumn{1}{c|}{HW} & \multicolumn{1}{c|}{Hybrid}   & \multicolumn{1}{c|}{Hybrid}  & \multicolumn{1}{c|}{Hybrid}  & \multicolumn{1}{c}{Hybrid} \\
		\multicolumn{1}{l|}{Secure Communication}         & \multicolumn{1}{c|}{yes}  & \multicolumn{1}{c|}{no} & \multicolumn{1}{c|}{no} &  \multicolumn{1}{c|}{yes} & \multicolumn{1}{c|}{yes}  & \multicolumn{1}{c|}{yes}  & \multicolumn{1}{c|}{no} & \multicolumn{1}{c}{no} \\
		\multicolumn{1}{l|}{Secure boot}         & \multicolumn{1}{c|}{yes}  & \multicolumn{1}{c|}{no} & \multicolumn{1}{c|}{no} &  \multicolumn{1}{c|}{yes} & \multicolumn{1}{c|}{yes}  & \multicolumn{1}{c|}{yes}  & \multicolumn{1}{c|}{no} & \multicolumn{1}{c}{no} \\
		\multicolumn{1}{l|}{Remote Attestation}         & \multicolumn{1}{c|}{yes}  & \multicolumn{1}{c|}{no} & \multicolumn{1}{c|}{no} & \multicolumn{1}{c|}{no} & \multicolumn{1}{c|}{yes} & \multicolumn{1}{c|}{no} & \multicolumn{1}{c|}{yes} & \multicolumn{1}{c}{no}  \\
		\multicolumn{1}{l|}{Malicious Code Modification Attacks Detection}         & \multicolumn{1}{c|}{yes}  & \multicolumn{1}{c|}{yes} & \multicolumn{1}{c|}{yes} &  \multicolumn{1}{c|}{yes} & \multicolumn{1}{c|}{yes} & \multicolumn{1}{c|}{yes}  & \multicolumn{1}{c|}{yes} & \multicolumn{1}{c}{yes}  \\
		\multicolumn{1}{l|}{Malicious Code Modification Attacks Protection}         & \multicolumn{1}{c|}{yes}  & \multicolumn{1}{c|}{no} & \multicolumn{1}{c|}{yes} &  \multicolumn{1}{c|}{yes} & \multicolumn{1}{c|}{yes} & \multicolumn{1}{c|}{yes}  & \multicolumn{1}{c|}{yes} & \multicolumn{1}{c}{yes}  \\ 
		\multicolumn{1}{l|}{Recovery from Malicious Code Modification Attacks}         & \multicolumn{1}{c|}{yes}  & \multicolumn{1}{c|}{yes} & \multicolumn{1}{c|}{yes} &  \multicolumn{1}{c|}{partial} & \multicolumn{1}{c|}{no}  & \multicolumn{1}{c|}{no}  & \multicolumn{1}{c|}{no} & \multicolumn{1}{c}{no}  \\
		\multicolumn{1}{l|}{Formal Verification of Hardware}         & \multicolumn{1}{c|}{no}  & \multicolumn{1}{c|}{no} & \multicolumn{1}{c|}{no} &  \multicolumn{1}{c|}{no} & \multicolumn{1}{c|}{no} & \multicolumn{1}{c|}{yes} & \multicolumn{1}{c|}{yes} & \multicolumn{1}{c}{yes} \\   
		\multicolumn{1}{l|}{Formal Verification of  Software}         & \multicolumn{1}{c|}{no}  & \multicolumn{1}{c|}{no} & \multicolumn{1}{c|}{no} &  \multicolumn{1}{c|}{no} & \multicolumn{1}{c|}{no} & \multicolumn{1}{c|}{yes} & \multicolumn{1}{c|}{yes} & \multicolumn{1}{c}{yes} \\  
		\multicolumn{1}{l|}{Formal hardware/software co-Verification}         & \multicolumn{1}{c|}{no}  & \multicolumn{1}{c|}{no} & \multicolumn{1}{c|}{no} &  \multicolumn{1}{c|}{no} & \multicolumn{1}{c|}{no} & \multicolumn{1}{c|}{yes} & \multicolumn{1}{c|}{yes} &
		\multicolumn{1}{c}{yes} \\  
		\bottomrule
		
	\end{tabular}
\end{table*}\label{quantfv}
\begin{table}[H]
	\caption{Security Properties verification results on i7-NUC @3.4GHz}\label{vft}
	\begin{tabular}{@{}lcccc@{}} 
		\hline
		\multicolumn{1}{l|}{Security Properties Specification}        & \multicolumn{1}{c|}{\emph{Time (s)}} &	\multicolumn{1}{c|}{ Hw } &  \multicolumn{1}{c|}{Fw} & \multicolumn{1}{c}{Results}\\ \hline 
		\multicolumn{1}{l|}{Start-up Checking}    & \multicolumn{1}{c|}{0.02}  & \multicolumn{1}{c|}{yes} & \multicolumn{1}{c|}{no}  &\multicolumn{1}{c}{{\includegraphics[width=.3cm,height=.3cm]{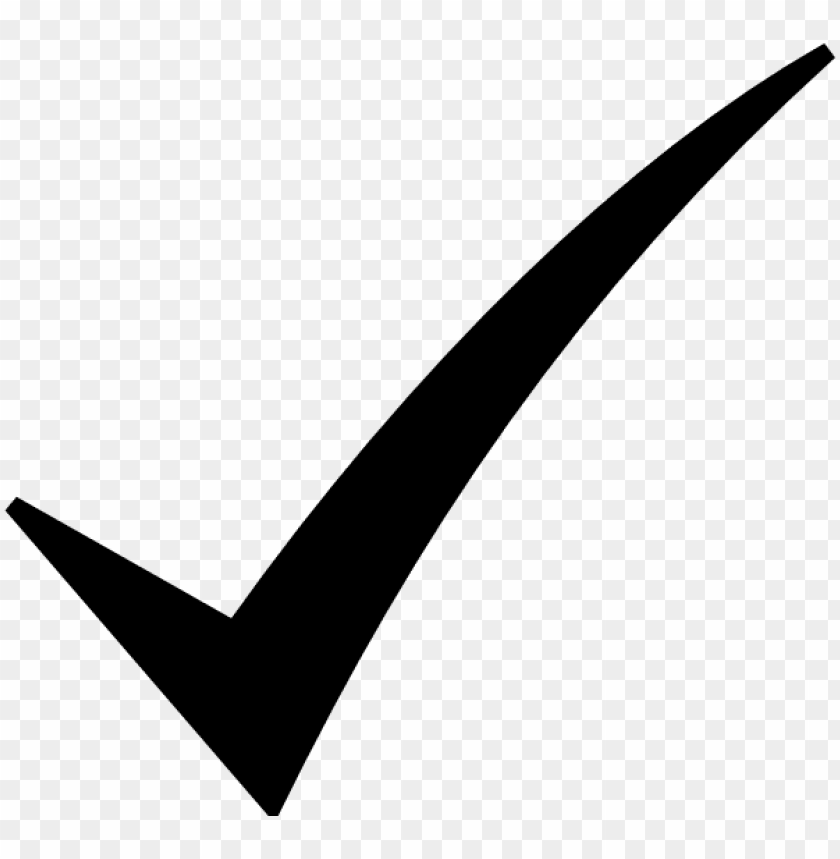}}}   \\
		\multicolumn{1}{l|}{Peripheral Initialization}      & \multicolumn{1}{c|}{0.03}  & \multicolumn{1}{c|}{yes} & \multicolumn{1}{c|}{yes}  &\multicolumn{1}{c}{{\includegraphics[width=.3cm,height=.3cm]{figs/right}}}   \\
		\multicolumn{1}{l|}{Key Confidentiality}         & \multicolumn{1}{c|}{0.02}  & \multicolumn{1}{c|}{yes} & \multicolumn{1}{c|}{no} &  \multicolumn{1}{c}{{\includegraphics[width=.3cm,height=.3cm]{figs/right}}} \\
		\multicolumn{1}{l|}{Access Control Enforcement}  & \multicolumn{1}{c|}{0.03}  & \multicolumn{1}{c|}{yes} & \multicolumn{1}{c|}{yes} &   \multicolumn{1}{c}{{\includegraphics[width=.3cm,height=.3cm]{figs/right}}}  \\ 
		\multicolumn{1}{l|}{Controlled Invocation} & \multicolumn{1}{c|}{0.02}  & \multicolumn{1}{c|}{yes} & \multicolumn{1}{c|}{no}  & \multicolumn{1}{c}{{\includegraphics[width=.3cm,height=.3cm]{figs/right}}} \\ 
		\multicolumn{1}{l|}{Attack Detection}  & \multicolumn{1}{c|}{0.02}  & \multicolumn{1}{c|}{yes} & \multicolumn{1}{c|}{yes}  & \multicolumn{1}{c}{{\includegraphics[width=.3cm,height=.3cm]{figs/right}}}  \\     
		\multicolumn{1}{l|}{Correct Frame Locations}   & \multicolumn{1}{c|}{0.02}  & \multicolumn{1}{c|}{no} & \multicolumn{1}{c|}{yes} & \multicolumn{1}{c}{{\includegraphics[width=.3cm,height=.3cm]{figs/right}}} \\
		\multicolumn{1}{l|}{Validate Frame Size}       & \multicolumn{1}{c|}{0.02}  & \multicolumn{1}{c|}{no} & \multicolumn{1}{c|}{yes}  & \multicolumn{1}{c}{{\includegraphics[width=.3cm,height=.3cm]{figs/right}}} \\ 
		\multicolumn{1}{l|}{Functional Correctness}      & \multicolumn{1}{c|}{0.2}  & \multicolumn{1}{c|}{yes} & \multicolumn{1}{c|}{yes}  &\multicolumn{1}{c}{{\includegraphics[width=.3cm,height=.3cm]{figs/right}}}   \\
		\hline
	\end{tabular}
\end{table}

Note that FVCARE verifies the crypto-core's functional correctness using hardware verification as discussed in subsection~\ref{hwfv}. Furthermore, the system's functional correctness of secure boot and RA features is validated using the hardware approach's software abstraction. The formal verification of the secure boot for the test application of 5.6~KB takes 0.2 seconds. 
\subsection{\bf{Comparison with the state-of-the-art solutions}}
\par For the state-of-the-art comparison, FVCARE has identified several recent implementations of secure boot and RA techniques as listed in Table~II. As can be seen from Tabel II majority of the available, secure boot and RA implementations focus on detecting and preventing malicious code modification attacks. However, it lacks protection from attacks and mostly restarts the system or leaves it in a non-operational state. Recent implementations Healed and \cite{Secerase:2010} provides recovery, but they do not have secure boot, RA, and formal verification support. Other implementation \cite{Vrased:2019} shows formal verification of RA module using LTL specification, with two unsuitable design choices: 1) it uses software-based crypto-core (HACL*) for digest computation in RA, 2) it does not have secure boot support, and 3) it recommends systems reset to prevent the attacks. Work presented by \cite{fvap:2016} uses the instruction-level abstraction of hardware approach for formal verification of security primitives such as secure boot. ILA approach is very restrictive, ISA specific, and limited to scale. Therefore, not suitable for scalable end-to-end system verification. Another work presented in \cite{ray:2019} demonstrates the use of the source-level abstraction of the hardware and bounded model checking for industry-standard SoC's security verification. Furthermore, Recent implementations CARE \cite{Care:2020} and SRACARE \cite{Sracare:2020} demonstrates resilient small embedded system design with secure boot, RA and on-board recovery techniques. For various security reasons and practical use-cases, our hypothesis required secure boot, RA, and onboard recovery such as CARE \cite{Care:2020}, and SRACARE \cite{Sracare:2020}. FVCARE uses the source-level abstraction of the hardware approach and enhances the model checking capabilities by using the Frama-C tool with different K-induction plugins, Weakness prediction, assertion, and bounded model checking using LTL. Thus, FVCARE is the first implementation that integrates hardware and software verification methods and demonstrates the end-to-end co-verification technique for SRACARE based systems with secure boot, RA, and onboard recovery mechanisms.   
\section{Conclusion}
FVCARE  provides the end-to-end co-verification framework for SRACARE based systems with secure boot, RA, and onboard recovery. It uses the abstraction of hardware as a software technique for formal verification of design specifications, security properties, and the system's functional correctness. Also, it uses hardware verification techniques for verification of initial system and registers settings, access control policies. It demonstrates formal verification of hardware using Linear Temporal Logic (LTL) properties and using model checking NuSMV tool. FVCARE leverages the software abstraction of the hardware approach for software verification and uses a novel Frama-C framework with different plugins for formal verification of the system represented as software abstraction of the hardware. FVCARE demonstrates the first practical implementation of a formal co-verification framework.

\bibliographystyle{IEEEtran}

\bibliography{references}

%
\IEEEpeerreviewmaketitle

%

\begin{IEEEbiography}[{\includegraphics[width=1in,height=1.25in,clip,keepaspectratio]{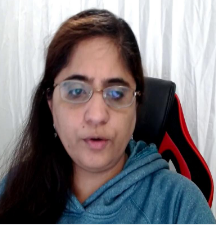}}]{Avani Dave PhD. Candidate, CSEE, University of Maryland Baltimore County, MD, USA.}
\end{IEEEbiography}
\vspace{-15pt}
\begin{IEEEbiography}[{\includegraphics[width=1in,height=1.25in,clip,keepaspectratio]{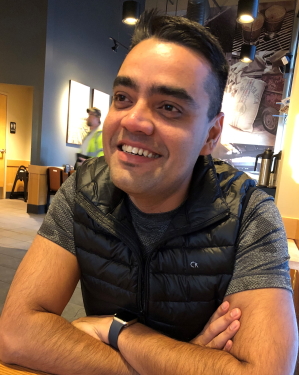}}] {Nilanjan Banerjee Professor, CSEE, University of Maryland Baltimore County, MD, USA.}
\end{IEEEbiography}\vspace{-2em}
\begin{IEEEbiography}[{\includegraphics[width=1in,height=1.25in,clip,keepaspectratio]{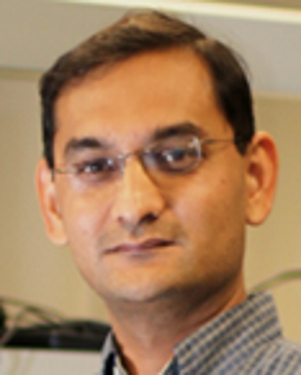}}]{Chintan Patel \IEEEmembership{Member,~IEEE,} Associate Professor, CSEE, University of Maryland Baltimore County, MD, USA.}
\end{IEEEbiography}




\end{document}